# Emotion-Recognition Using Smart Watch Sensor Data: Mixed-Design Study

Juan C Quiroz, PhD
Department of Computer Science and Engineering
University of Nevada, Reno
Reno, NV, USA
quiroz@cse.unr.edu

Elena Geangu, PhD
Psychology Department
University of York
Heslington, York, YO10 5DD, UK
elena.geangu@york.ac.uk

Min Hooi Yong, PhD
Department of Psychology
Sunway University
Bandar Sunway, Malaysia
mhyong@sunway.edu.my

## Abstract

**Background:** Research in psychology has shown that the way a person walks reflects that person's current mood (or emotional state). Recent studies have used smartphones to detect emotional states from movement data.

**Objective:** This study investigates the use of movement sensor data from a smart watch to infer an individual's emotional state. We present our findings on a user study with 50 participants.

**Methods:** The experimental design is a mixed-design study; within-subjects (emotions; happy, sad, neutral) and between-subjects (stimulus type: audio-visual "movie clips", audio "music clips"). Each participant experienced both emotions in a single stimulus type. All participants walked 250m while wearing a smart watch on one wrist and a heart rate monitor strap on their chest. They also had to answer a short questionnaire (20 items; PANAS) before and after experiencing each emotion. The heart rate monitor served as supplementary information to our data. We performed time-series analysis on the data from the smart watch and a t-test on the questionnaire items to measure the change in emotional state. The heart rate data was analyzed using one-way ANOVA. We extracted features from the time-series using sliding windows and used the features to train and validate classifiers that determined an individual's emotion.

**Results:** We had 50 young adults participate in our study, with 49 included for the affective PANAS questionnaire and 44 for the feature extraction and building of personal models. Participants reported feeling less negative affect after watching sad videos or after listening to the sad music, P < .006. For the task of emotion recognition using classifiers, our results show that the personal models outperformed personal baselines, and achieved median accuracies higher than 78% for all conditions of the design study for the binary classification of happiness vs sadness.

**Conclusions:** Our findings show that we are able to detect the changes in emotional state with data obtained from the smartwatch as well as behavioral responses. Together with the high accuracies achieved across all users for the classification of happy vs sad emotional states, this is further evidence for the hypothesis that movement sensor data can be used for emotion recognition.

## Introduction

Our emotional state is often expressed in a variety of means, such as face, voice, body posture, and walking gait (1,2). Many studies are conducted in strict laboratory settings, which may impede the overall ecological validity of the findings. Having a strong ecological validity is important because emotional expression or display in any modality is not entirely dependent on conscious action or function. Instead, emotional expressions are essentially a response to a particular affective stimulus/experience and this response might be reduced as a result of social desirability in a lab.

Speech, video, and physiological data have been analyzed to determine the emotional state of a person (3,4), but these analyses usually rely on recordings performed in laboratory environments with limited ecological validity. In order to formulate theoretical models of emotions and affective health which take into account the richness of everyday life, we need to measure affective states unobtrusively during these situations. Smartphones include sensors, such as accelerometers, which have the potential to be sensitive to changes in people's affective states, and thus could provide rich and accessible information in this respect. For example, we know that the way we walk reflects whether we feel happy or sad (2). This paper analyzes movement sensor data recorded via a smart watch in relation to changes in emotions.

Prior work on emotion detection from smartphone data includes the analysis of typing behavior on a smartphone (5,6) and smartphone usage (7,8). The EmotionSense system performed emotion detection directly on mobile phones via the analysis of speech, with additional sensors collecting information about the user and his environment (9). There are however some indications that movement sensor data collected by smartphones could be a viable solution for inferring emotion, as opposed to inferring movement or physical activities. Cui et al. attempted to record participants' movement with smartphones strapped to their ankles and wrists, thus impairing ecological validity (10). Happiness and anger were elicited with video stimuli and emotional state classifiers were trained with accelerometer data (10). Zhang et al. also focused on happiness and anger, but they recorded the movement data with smart bracelets (11). The accuracies in detecting these emotions ranged from 60% to 91.3% across all subjects using 10-fold cross-validation (11).

These cases have motivated further research on the tracking and analysis of sensor data from smartphones and wearables with the goal of monitoring and intervention for patients suffering from mental illnesses or substance abuse (12,13). Further validation is needed for the hypothesis that movement sensor data can be used to recognize emotional states. Movement data is of particular interest because accelerometers and gyroscopes are standard sensors in smartphones, wearables, and fitness trackers. Movement data collection is unobtrusive and it requires no user input (14), which gives us reliable data in the real world without adding the possibility of having social desirability responses.

Toward this end, we make the following contributions. First, we conducted a mixed-design study (Fig. 1) with 50 participants to test two types of stimuli for eliciting emotional responses from participants: audio-visual and audio. Participants wore a smart watch on their wrist and a heart rate strap on their chest. The heart rate strap was included to supplement the data collected from the Positive Affect and Negative Affect (PANAS) scores (15). After or while watching the emotional stimuli, participants walked, and the process was repeated three times, for each type of emotion: happy, neutral, and sad. We extracted features from the sensor data and built classifiers (personal models) that recognized the emotional state of the user. Our results show that the personal models outperformed personal baselines, and achieve median accuracies higher than 78% for all conditions of the design study for the binary classification of happiness vs sadness. This paper is an extended version of preliminary findings published in (16).

## Methods

### Participants

Fifty young adults participated in this study (43 females, $M$ = 23.18 years, $SD$ = 4.87). All participants were recruited in a university campus (North-West UK) via announcements on notice boards and word of mouth. Each participant was given £7 for participation. None of the participants reported any vision or hearing difficulties and could walk unassisted.

We obtained ethical approval from Sunway University Ethics Board (SUREC 2016/05) and had it validated by Lancaster University to conduct both validation and actual main study experiment.

## Materials

The study included two types of stimuli: a) audio-visual and b) audio.

### Audio-visual

Audio-visual clips were selected from commercial movies with the potential of being perceived as having emotional meaning (i.e., sadness and happiness) and to elicit emotional responses. The commercial movies were selected from Gross and Levenson (17), Bartollini (18), Schaefer et al. (19), and from five young adults (4 females, $M$ = 21.50 years). Another five participants (3 females, $M$ = 22.80 years, $SD$ = 1.30) were asked to identify each of these clips in terms of the emotion they felt while watching it, and the intensity of the emotion they felt using a 0-to-10 Likert scale (0: hardly, 10 – very much likely). They were also asked if they had watched that movie before. On average, only one participant had seen that movie before. The participants reported that they felt the emotion intended for all clips (100% accuracy) and the intensity ranged between 5.0 to 6.5 for happy and sad clips respectively. See Table 1 for the movie clips used in our study.

Table 1. Movies used to induce happy and sad emotions.

|  | Movie |
|---|---|
|  |  |
| **Happy** |  |
|  | 10 Things I Hate About You (1999) <br> Patrick serenades Katarina in stadium |
|  | When Harry Met Sally (1989) <br> Discussion of orgasms in cafe |
|  | There's Something about Mary (1998) <br> Mary hair gel scene |
|  | Monty Python (1975) <br> Black Knight fights King Arthur |
|  | Modern Times (1936) <br> Factory worker in assembly line |
|  | Love Actually (2003) <br> Arrival halls scene in Heathrow airport |
|  | Wall-E (2008) <br> EVA kisses Wall-E |
|  | Benny & Joon (1993) <br> Sam roll dance in diner |
|  |  |

| Sad | |
|---|---|
| | Interstellar (2014)<br>Cooper watches video messages sent by his children |
| | Click (2006)<br>Michael rewinds his past to recall for not saying goodbye to his father |
| | Hachi (2009)<br>Hachiko waits at the train station |
| | Shawshank Redemption (1994)<br>Death of Brooks |
| | Saving Private Ryan (1998)<br>Mother is informed of the deaths of all of Private Ryan's brothers |
| | Marley & Me (2008)<br>Marley is euthanized in the veterinarian clinic |
| | The Champ (1979)<br>Boy cries at father's death |
| | My Girl (1991)<br>Thomas' funeral |

## Audio

For the audio stimuli, pieces of classical music were chosen known to elicit happy, sad, and emotionally neutral states as reported by (20). See Table 2 for selected clips.

Table 2. Musical pieces used to induce happy and sad emotions.

| | Piece | Composer |
|---|---|---|
| **Happy** | | |
| | Carmen: Chanson du toreador | Bizet |
| | Allegro—A little night music | Mozart |
| | Rondo allegro—A little night music | Mozart |
| | Blue Danube | Strauss |
| | Radetzky march | Strauss |
| | | |
| **Sad** | | |
| | Adagio in sol minor | Albinoni |
| | Kol Nidrei | Bruch |
| | Solveig's song – Peer Gynt | Grieg |
| | Concerto de Aranjuez | Rodrigo |
| | Suite for violin & orchestra A minor | Sinding |
| | | |
| **Neutral** | | |
| | L'oiseau prophete | Schumann |
| | Claire de lune | Beethoven |
| | Claire de lune | Debussy |
| | Symphony no. 2 C minor | Mahler |

|  | La traviata | Verdi |
|--|-------------|-------|
|  | Pictures at an exhibition | Mussorgsky |
|  | Water music—passepied | Handel |
|  | Violin romance no. 2 F major | Beethoven |
|  | Water music—minuet | Handel |
|  | The planets—Venus | Holst |
|  | L'oiseau prophete | Schumann |

## Procedure

All participants were presented with happy, sad, and neutral stimuli. A third of the participants (N = 18) were presented with the audio-visual stimuli (i.e., videos), while the other remaining participants (N = 32) with the audio stimuli (i.e., classical music). Half of the participants (N = 16) who were assigned to the audio stimuli, listened to them prior to walking, while the other half (N = 16) listened to the stimuli while they were walking. Eighteen participants (N = 18) who were assigned to the emotional videos watched them prior to walking. Assignment to each condition was random. To counter the possible order effects, half of the participants had the sad stimuli first while the other half had the happy stimuli first. Each participant was tested individually, and the task took approximately 20 minutes to complete. All data was collected between 17:00 and 19:00 h to account for peak foot traffic.

The three conditions of the mixed-design study are presented in (Figure 1): (1) Condition 1 – watching the movie clip prior to walking; (2) Condition 2 – listening to the music prior to walking; and (3) Condition 3 – listening to music while walking.

Each participant was first greeted by the experimenter at one end of the corridor and proceeded to put on various items. First, the participant had the heart rate sensor (Polar H7) strapped snugly around their chest. The corresponding watch (Polar M400) was strapped onto the experimenter's wrist. The watch was set to "other indoor" sport profile. Second, the participant strapped a smart watch (Samsung Gear 2) on their left wrist. Participants wore the sensors for the entire duration of the experiment. The smart watch included a tri-axial accelerometer and a tri-axial gyroscope. The sampling rate of the smart watch is advertised as 25 Hz, but our results show that actual sampling rate on average was 23.8 Hz. For the smart watch, we developed a Tizen application that recorded accelerometer and gyroscope sensor data.

The participant rated their current mood state using Positive and Negative Affect Schedule (PANAS) (21) on a 7" tablet. PANAS contains ten adjectives for positive (e.g. joy) and negative feelings (e.g. anxiety) respectively. Scores can range from 10–50, with higher scores representing higher levels of affect. The heart rate sensor was used in the study to supplement the data collected from the PANAS scores (15).

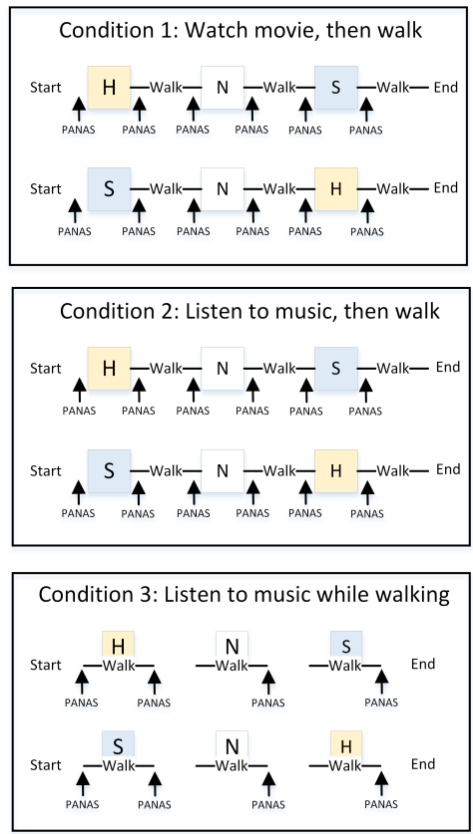

**Figure 1**: Mixed-design study with three conditions. The three conditions were used to determine the stimulus that would better induce the target emotional states on participants.

For Conditions 1 and 2, in which the stimulus presentation occurred before walking, participants placed a pair of headphones to listen or watch the assigned stimuli (e.g. sad music or happy movie) while located at the start of a walking route. At the end of the stimulus, the participant walked to the end of the route and back to the starting point. Participants were reminded not to make any stops in between. The route was represented by a 250m S-shaped corridor located on the ground floor of a university building. The experimenter followed the participant at a 125m distance discreetly to observe behavior and to ensure that heart rate monitoring was captured by the watch. Upon their return, the participant rated their mood using the same PANAS scales. Because of the initial mood induction, we always had a neutral condition in between happy and sad conditions to allow return to the baseline calm state. For all participants, the neutral stimulus was classical music for the audio type or a movie with classical music playing in the background depicting an everyday scene. The same procedure above - rating their initial mood using PANAS, watching or listening to a stimulus, walking along the corridor and back, and rating their mood – was applied to the neutral and second emotion.

In Condition 3 – listening while walking – the procedure was similar to the above except that the participant was listening to the assigned music while walking, and participants reported PANAS after walking.

## Feature Extraction

During the experiment, the experimenter recorded the time when each participant started and stopped walking. These times were used to identify the sensor data that corresponded to actual walking time. We discarded the sensor data during the time when participants were briefed and when participants watched/listened to the stimulus prior to walking.

The walking times were labeled according to the corresponding emotional stimulus presented before walking. For example, if the participant viewed a movie clip known to induce happiness, all of the features extracted from the subsequent walking data were labeled as happy. These labels were used to train classifiers for the recognition of happiness vs sadness. We present the classifier results for the two-class problem of detecting happy vs sad emotions and for the three-class problem of detection happy vs sad vs neutral emotions.

We first filtered the raw accelerometer data with a mean filter (window=3). Features were extracted from sliding windows with a size of one second (24 samples) with 50% overlap. Our feature extraction approach is similar to that used for activity recognition from smartphone accelerometer data (22,23). That is, each window is treated as an independent sample (feature vector). We address neighborhood bias when building models from accelerometer sliding windows in the results section (24).

For each window of the tri-axial accelerometer and tri-axial gyroscope data, we extract 17 features (23): (1) mean, (2) standard deviation, (3) max, (4) min, (5) energy, (6) kurtosis, (7) skewness, (8) root mean square, (9) root sum square, (10) sum, (11) sum of absolute values, (12) mean of absolute values, (13) range, (14) median, (15) upper quartile, (16) lower quartile, and (17) median absolute deviation. These 17 features were extracted from each of the three axes of the accelerometer data and each of the three axes of the gyroscope data, resulting in 102 features. We also calculated the angle between the signal mean (within a window) and the x-axis, y-axis, and z-axis (three features); the standard deviation of the signal magnitude (one feature); and the heart rate (one feature), for a total of 107 features for the feature vector of a window. Unless otherwise stated, we used all 107 features for classification. However, we do explore classification performance based on features corresponding to certain sensors: (1) accelerometer, gyroscope, and heart rate; (2) accelerometer and heart rate; and (3) accelerometer.

We divided the data by condition and built personal models with the features extracted from each window (25). In personal models, the training and testing data comes from a single user. In our case, we built 44 personal models (the data from 6 participants was discarded because of missing data and other recording errors), with each model evaluated using stratified 10-fold cross-validation repeated 10 times. For each participant, we had an average of 403.29 (SD=55.62) samples labeled as happy, 403.67 (SD=51.46) samples labeled as sad, and 402.93 (SD=50.24) samples labeled as neutral. Out of the 44 personal models built, 16 were from Condition 1 (watch movie and then walk), 14 were from Condition 2 (listen to music and then walk), and 14 were from Condition 3 (listen to music and then walk).

We compared random forests—with 100 estimators (RF), logistic regression—with L2 regularization (LR), and a baseline classifier that picked the majority class as the prediction (BL). The python scikit-learn library was used for the training and testing of these classifiers. Since the number of samples labeled as happy vs sad for each participant were approximately the same, the baseline classifier predicted each window as happy vs sad with about a 50% probability (i.e. all samples for user *i* were classified as happy, resulting in about a 50% accuracy). For binary classification of happy vs sad, we use the accuracy, the F1 score, and the area under the receiver operating characteristic curve (ROC AUC) to assess classification performance. For multiclass classification of happy vs sad vs neutral, we use the accuracy and the F1 score.

## Results

### Ecological validity checks
When asked about their experience in using a smart gadget, most of the participants were familiar and comfortable with the smart watch, but not the Polar heart rate monitor. They did not notice anything unusual about the study, which may have influenced their walking gait and behavioral response.

### Behavioral response to stimuli (PANAS outcomes)
We analyzed the PANAS responses for all conditions on the happy versus sad stimuli. One participant's data was excluded for being incomplete, thus leaving 49 for analyses (15 for Condition 1, 18 for Condition 2, 16 for Condition 3). We first reviewed the normality and found that the data was normally distributed for Conditions 1 and 2, but not for Condition 3 (visual histograms were skewed and Shapiro-Wilk *P*s < .01). See Multimedia Appendix 1 for PANAS scores for each emotion.

#### Condition 1: Watch movie and then walk
Participants reported a reduced negative affect after watching a sad movie clip (M = 14.94, SD = 6.79) compared to before (M = 19.00, SD = 7.20), t (16) = 3.16, *P* = .006. There was no significant difference for positive affect for the sad movie, t (16) = .08, *P* = .94 and for both affect in the other two emotions (happiness and neutral), all *P*s > .10.

#### Condition 2: Listen to music and then walk
For the sad music, participants reported an increased positive affect after the walk (M = 24.00, SD = 5.33) compared to prior (M = 20.31, SD = 5.79), t (15) = 2.96, *P* = .01, and a reduced negative affect after (M = 11.69, SD = 3.34) as to before (M =

13.63, SD = 5.12), t (15) = 2.78, *P* = .014. Participants reported a reduced positive affect after listening to happy music (M = 26.38, SD = 6.96) compared to before (M = 29.56, SD = 5.17), t (15) = 2.62, *P* = .02, but no significant difference for the negative affect, t (15) = 1.60, *P* = .13. There is no significant difference for the neutral music for both affect, both *P*s > .76.

### *Condition 3: Listen to music while walking*
Participants reported an increased negative affect when walking listening to happy (M=13.31, SD = 4.88) compared to neutral (M=15.00, SD = 5.44) music, Z = 2.64, *P* = .08. No other significant differences were observed, all *P*s > .13.

### *Heart rate*
We planned to verify the data obtained from PANAS and determine whether our participants experienced an accelerated or decelerated heart rate as a result of the emotional stimuli (15). From the 50 participants, we had some data loss due to technical fault (n= 9; 3 from Condition 1, 3 from Condition 2, 3 from Condition 3), thus leaving us with 41 participants' data. We first reviewed the descriptive statistics and find that the data is normally distributed. A one-way between-subjects analysis of variance ANOVA was conducted to compare the effect of emotion (happy, sad, neutral) on participants' heart rate. We did not find any significant effect of emotion on their heart rate at the *P* <.05 level for the three conditions [F(2, 120) = 0.13, *P* = 0.88] (see Table 3 for means and standard deviation).

Table 3. Mean heart rate and standard deviations in brackets for all three emotion conditions.

| Happy | Sad | Neutral |
|---|---|---|
| 104.43 (14.55) | 91.68 (16.31) | 105.77 (14.50) |

### Emotion Recognition

### *Happy vs Sad*
Figure 2 illustrates three boxplot sets, one for each condition, showing the distribution of accuracies for the personal model of each participant. For all three conditions, the personal baselines have accuracies in the range of 50% to 54%. For all conditions, both random forests and logistic regression outperform the baseline, with the accuracies being in the range of 62 to 99%. Condition 1 (movie) and condition 3 (music while walking) resulted in the highest classification accuracies, with median accuracies over 82%.

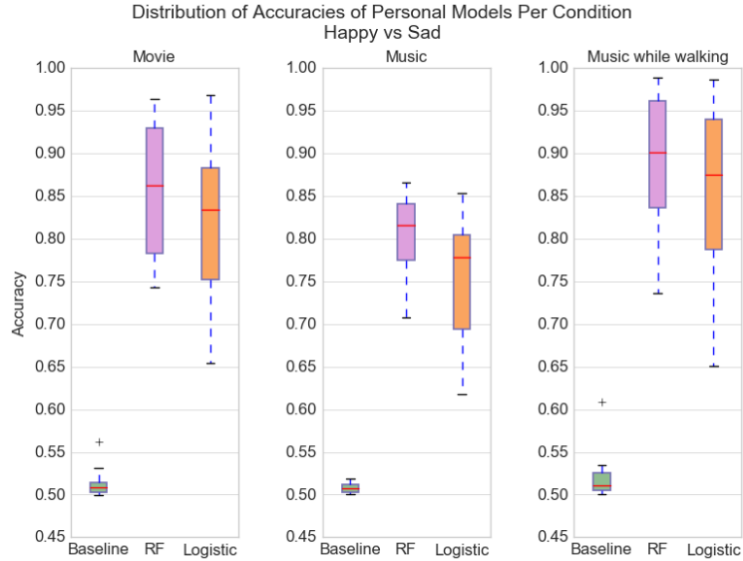

**Figure 2**: Boxplot of classification accuracies for participants divided by conditions. Algorithms tested were baseline (pick majority), random forests, and logistic regression. Outliers are indicated by +. The highest classification accuracies were achieved with Condition 1 (movie) and Condition 3 (music while walking). For all conditions, the models achieved accuracies greater than 78% for over half the users.

We used the user lift framework to quantify whether a personal model is better than a personal baseline for each user [26]. We calculated the user lift as the difference in accuracy of the personal classifier and the personal baseline (classifier accuracy – personal baseline accuracy). We used the nonparametric permutation test to determine whether the user lifts had a mean greater than 0 (see Table 4). Figure 3 shows the calculated user lift for each participant using random forests and logistic regression. We included this figure since average user lift can obscure the presence of negative user lift for some participants.

Table 4. Average user lift and average personal model accuracy per condition. Using features extracted from the accelerometer, gyroscope, and heart rate data results in the highest accuracies. Using only features from accelerometer data results in lower accuracies. Over all of the personal models, the average user lift is greater than 0 for all conditions, indicating that the trained personal models outperform the baseline.

| | Features | Model | AUC | F1 | Accuracy | User Lift | *P*-value |
|---|---|---|---|---|---|---|---|
| **Condition 1: Watch movie then walk** | | | | | | | |
| | Acc, Gyro, HR | BL | 0.500 (0.000) | 0.348 (0.017) | 0.513 (0.015) | | |
| | Acc, Gyro, HR | LR | 0.876 (0.085) | 0.817 (0.089) | 0.818 (0.089) | 0.305 | 0.000 |
| | Acc, Gyro, HR | RF | 0.923 (0.059) | 0.854 (0.073) | 0.854 (0.073) | 0.342 | 0.000 |
| **Condition 2: Listen to music then walk** | | | | | | | |
| | Acc, Gyro, HR | BL | 0.500 (0.000) | 0.342 (0.007) | 0.508 (0.006) | | |
| | Acc, Gyro, | LR | 0.812 | 0.748 | 0.748 | 0.240 | 0.000 |

| | | | | | | |
|---|---|---|---|---|---|---|
| HR | | (0.081) | (0.071) | (0.071) | | |
| Acc, Gyro, HR | RF | 0.887 (0.046) | 0.806 (0.047) | 0.806 (0.047) | 0.298 | 0.000 |
| **Condition 3: Listen to music while walking** | | | | | | |
| Acc, Gyro, HR | BL | 0.500 (0.000) | 0.356 (0.031) | 0.520 (0.027) | | |
| Acc, Gyro, HR | LR | 0.900 (0.096) | 0.849 (0.107) | 0.849 (0.107) | 0.329 | 0.000 |
| Acc, Gyro, HR | RF | 0.948 (0.057) | 0.890 (0.081) | 0.891 (0.080) | 0.371 | 0.000 |
| | | | | | | |
| **Condition 1: Watch movie then walk** | | | | | | |
| Acc, HR | BL | 0.500 (0.000) | 0.348 (0.017) | 0.513 (0.015) | | |
| Acc, HR | LR | 0.809 (0.105) | 0.752 (0.099) | 0.753 (0.099) | 0.240 | 0.000 |
| Acc, HR | RF | 0.891 (0.081) | 0.821 (0.090) | 0.822 (0.089) | 0.309 | 0.000 |
| **Condition 2: Listen to music then walk** | | | | | | |
| Acc, HR | BL | 0.500 (0.000) | 0.342 (0.007) | 0.508 (0.006) | | |
| Acc, HR | LR | 0.729 (0.070) | 0.674 (0.055) | 0.675 (0.055) | 0.167 | 0.000 |
| Acc, HR | RF | 0.847 (0.046) | 0.768 (0.045) | 0.769 (0.045) | 0.261 | 0.000 |
| **Condition 3: Listen to music while walking** | | | | | | |
| Acc, HR | BL | 0.500 (0.000) | 0.356 (0.031) | 0.520 (0.027) | | |
| Acc, HR | LR | 0.876 (0.095) | 0.821 (0.106) | 0.821 (0.106) | 0.301 | 0.000 |
| Acc, HR | RF | 0.933 (0.067) | 0.871 (0.088) | 0.871 (0.088) | 0.351 | 0.000 |
| | | | | | | |
| **Condition 1: Watch movie then walk** | | | | | | |
| Acc | BL | 0.500 (0.000) | 0.348 (0.017) | 0.513 (0.015) | | |
| Acc | LR | 0.786 (0.097) | 0.726 (0.089) | 0.727 (0.089) | 0.215 | 0.000 |
| Acc | RF | 0.847 (0.076) | 0.768 | 0.774 (0.077) | 0.261 | 0.000 |
| **Condition 2: Listen to music then walk** | | | | | | |
| Acc | BL | 0.500 (0.000) | 0.342 (0.007) | 0.508 (0.006) | | |
| Acc | LR | 0.708 (0.056) | 0.657 (0.047) | 0.658 (0.047) | 0.150 | 0.000 |
| Acc | RF | 0.783 (0.051) | 0.712 (0.042) | 0.713 (0.042) | 0.205 | 0.000 |
| **Condition 3: Listen to music while walking** | | | | | | |
| Acc | BL | 0.500 (0.000) | 0.356 (0.031) | 0.520 (0.027) | | |
| Acc | LR | 0.848 (0.086) | 0.789 (0.096) | 0.790 (0.095) | 0.269 | 0.000 |
| Acc | RF | 0.899 (0.066) | 0.825 (0.080) | 0.825 (0.079) | 0.305 | 0.000 |

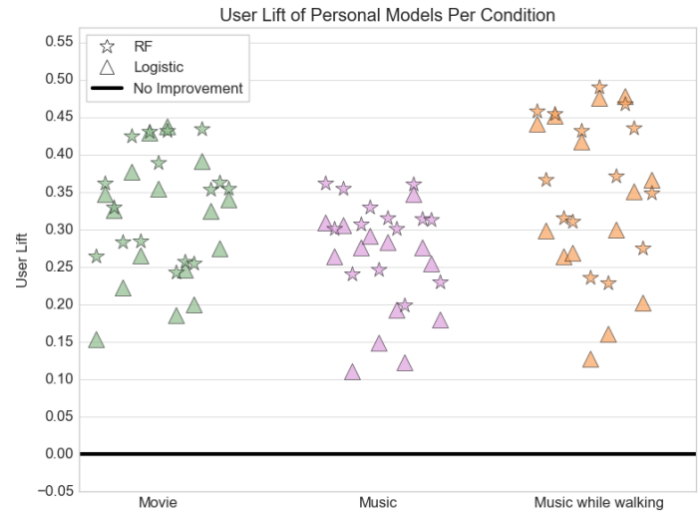

**Figure 3**: The user lift for personal models per condition. The random forest user lift is calculated as (random forest accuracy – baseline accuracy) and the logistic regression user lift is calculated as (logit accuracy – baseline accuracy). The personal models achieve higher accuracies than the personal baseline models.

### *Happy vs Neutral vs Sad*

Figure 4 shows the distribution of accuracies of personal models for the 3-class classification task of predicting happy-neutral-sad emotional states. We used all features (from the acceleration, angular velocity, and heart rate) for classification. While the personal models on average outperform the baseline, the accuracies are lower compared to the accuracies achieved when predicting only happy vs sad. As the number of samples for each class is approximately the same, the baseline predicting the majority class is only able to classify correctly about a third of the testing samples. See Table 5 for the user lift results.

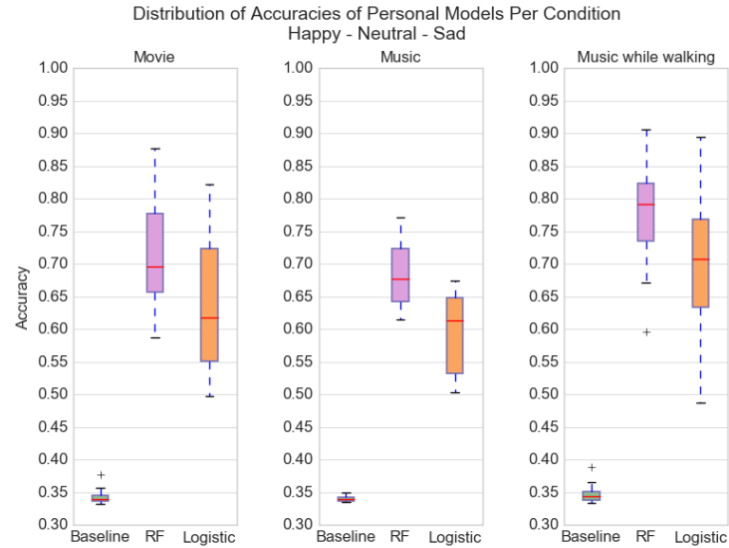

**Figure 4**: Classification accuracies for participants divided by conditions for the recognition of happiness, sadness, and neutral emotional states. The lower accuracies when recognizing the neutral emotional state indicates that the neutral walking data does have more similarities to the happy and sad walking data, which may indicate the need for additional features.

Table 5. Average user lift and average personal model accuracy per condition for the three-class classification task of predicting happy-neutral-sad. The personal models outperform the personal baselines, but overall accuracy is lower compared to the binary classification of happy vs sad.

| | Model | F1 | Accuracy | User Lift | *P*-value |
|---|---|---|---|---|---|
| **Condition 1: Watch movie then walk** | | | | | |
| | BL | 0.175 (0.010) | 0.343 (0.011) | | |
| | LR | 0.632 (0.103) | 0.635 (0.103) | 0.292 | 0.000 |
| | RF | 0.722 (0.090) | 0.723 (0.090) | 0.380 | 0.000 |
| **Condition 2: Listen to music then walk** | | | | | |
| | BL | 0.173 (0.004) | 0.340 (0.004) | | |
| | LR | 0.591 (0.062) | 0.594 (0.061) | 0.254 | 0.000 |
| | RF | 0.684 (0.048) | 0.685 (0.047) | 0.345 | 0.000 |
| **Condition 3: Listen to music while walking** | | | | | |
| | BL | 0.180 (0.014) | 0.348 (0.015) | | |
| | LR | 0.709 (0.113) | 0.711 (0.113) | 0.363 | 0.000 |
| | RF | 0.781 (0.087) | 0.782 (0.087) | 0.434 | 0.000 |

### Emotion cross-validation

We conducted an experiment to assess the effect of neighborhood bias in the evaluation of our models using random cross-validation. In this experiment, we conducted 10-fold cross-validation for each personal model, but the testing fold that was held out during each iteration held out a contiguous happy data block or a contiguous sad data block. The goal was to determine with higher confidence whether the classifiers were learning patterns associated with the emotions, as opposed to just learning to distinguish between different walking periods. In addition, this type of validation takes into consideration neighborhood bias, which can lead to overly optimistic performance estimates (24). The results (see Figure 5 and Table 6) show that accuracies across all conditions drop compared to accuracies when using random cross-validation. However, the performance of the models remains higher than personal baselines, with the exception of a few users. Only a quarter of the baseline models under Condition 1 and Condition 3 achieve accuracies between 0 and 0.5, the rest have accuracies of 0. This is expected, as a baseline model predicted on the majority class will achieve an accuracy of 0 when tested on a contiguous block of the opposite class.

We conclude that for at least half of the participants in Condition 1 (movie) and Condition 3 (music while walking) the models are likely learning patterns associated with the sad and happy emotions. In addition, the high accuracies indicate that the model performance is not a result of neighborhood bias (24).

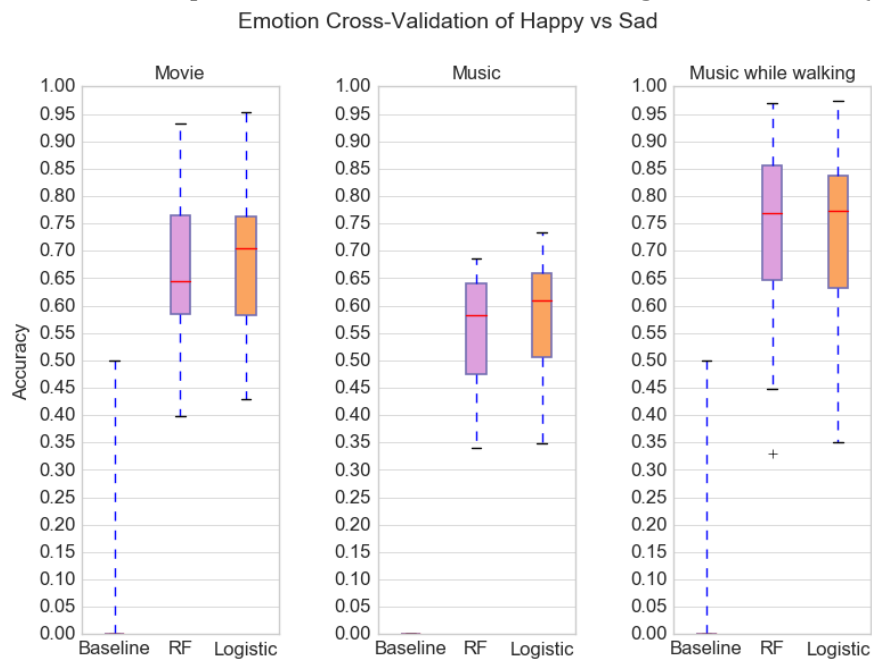

**Figure 5**: Boxplot of classification accuracies for participants divided by conditions. The results are for 10-fold cross-validation, with each fold in the training data consisting of contiguous windows from both happy and walking data, and the held-out test fold consisting of contiguous windows from either the happy or the sad walking data.

## Generalizing across users

We conducted leave-one-user-out cross-validation in order to assess how well a model trained on data from certain users would be able to generalize to a user for which no data is available. We compared both logistic regression and random forests. However, random forests performed similarly or worse than logistic regression, so we only discuss the results of the best performing logistic regression compared against the baseline (see Table 7). The low accuracies across all conditions shows that the behavior from user to user varies considerably, even when performing a similar action. Due to the small number of users per condition (<

18), the data may not be enough to make accurate predictions for users not included in the training set (24). However, it also highlights a limitation in our modeling approach, in that different features or more advanced models may be necessary in order to generalize across users. Ideally, deployment of an application should include an initial data collection and calibration phase, which can be used to build a high accuracy personal model for each user.

Table 6. Average user lift and average personal model accuracy per condition. The personal models outperform the personal baselines, but overall accuracy is poor.

| | Model | F1 | Accuracy | User Lift | $P$-value |
|---|---|---|---|---|---|
| **Condition 1: Watch movie then walk** | | | | | |
| | BL | 0.031 (0.121) | 0.031 (0.121) | | |
| | LR | 0.787 (0.104) | 0.682 (0.139) | 0.650 | 0.000 |
| | RF | 0.763 (0.112) | 0.651 (0.146) | 0.620 | 0.000 |
| **Condition 2: Listen to music then walk** | | | | | |
| | BL | 0.000 (0.000) | 0.000 (0.000) | | |
| | LR | 0.705 (0.099) | 0.575 (0.115) | 0.575 | 0.000 |
| | RF | 0.678 (0.105) | 0.543 (0.118) | 0.543 | 0.000 |
| **Condition 3: Listen to music while walking** | | | | | |
| | BL | 0.036 (0.129) | 0.036 (0.129) | | |
| | LR | 0.812 (0.140) | 0.723 (0.179) | 0.688 | 0.000 |
| | RF | 0.815 (0.148) | 0.731 (0.185) | 0.695 | 0.000k |

Table 7. Accuracy scores for leave-one-user-out cross-validation. Logistic regression performs poorly across all conditions, showing that using the data from different users to do emotion recognition on a different user is not possible with the current features and logistic regression.

| Model | $AUC$ | F1 | Accuracy |
|---|---|---|---|
| **Condition 1: Watch movie then walk** | | | |
| BL | 0.500 (0.000) | 0.342 (0.021) | 0.508 (0.018) |
| LR | 0.539 (0.137) | 0.461 (0.112) | 0.515 (0.090) |
| **Condition 2: Listen to music then walk** | | | |
| BL | 0.500 (0.000) | 0.332 (0.011) | 0.499 (0.010) |
| LR | 0.539 (0.084) | 0.467 (0.061) | 0.519 (0.059) |
| **Condition 3: Listen to music while walking** | | | |
| BL | 0.500 (0.000) | 0.323 (0.034) | 0.490 (0.032) |
| LR | 0.510 (0.173) | 0.476 (0.092) | 0.505 (0.082) |

## Model interpretability

We address model interpretability, that is, how the models are able to differentiate between emotions, by examining the information gain of the features. Random forests can be interpreted by examining feature importances and logistic regression can be interpreted by the sign and value of the coefficients. Random forests outperformed logistic regression in our results, so we limit our analysis to feature importances of random forests.

Because we are building personal models, features that might be important for one user may be less important for a different user. To show this, we plotted the distribution of feature importance values for each feature across all users using boxplots (See Figure 6). The boxplots are sorted by median and we include only the top 30 features for visibility, with the trend of the remaining features being about the same. To obtain the feature importances for each user, we computed the mean feature importance for each feature in the cross-validation folds, and divided each feature by the maximum feature importance value. Thus, a value of 1.0 indicates that a feature was the most important amongst all the features.

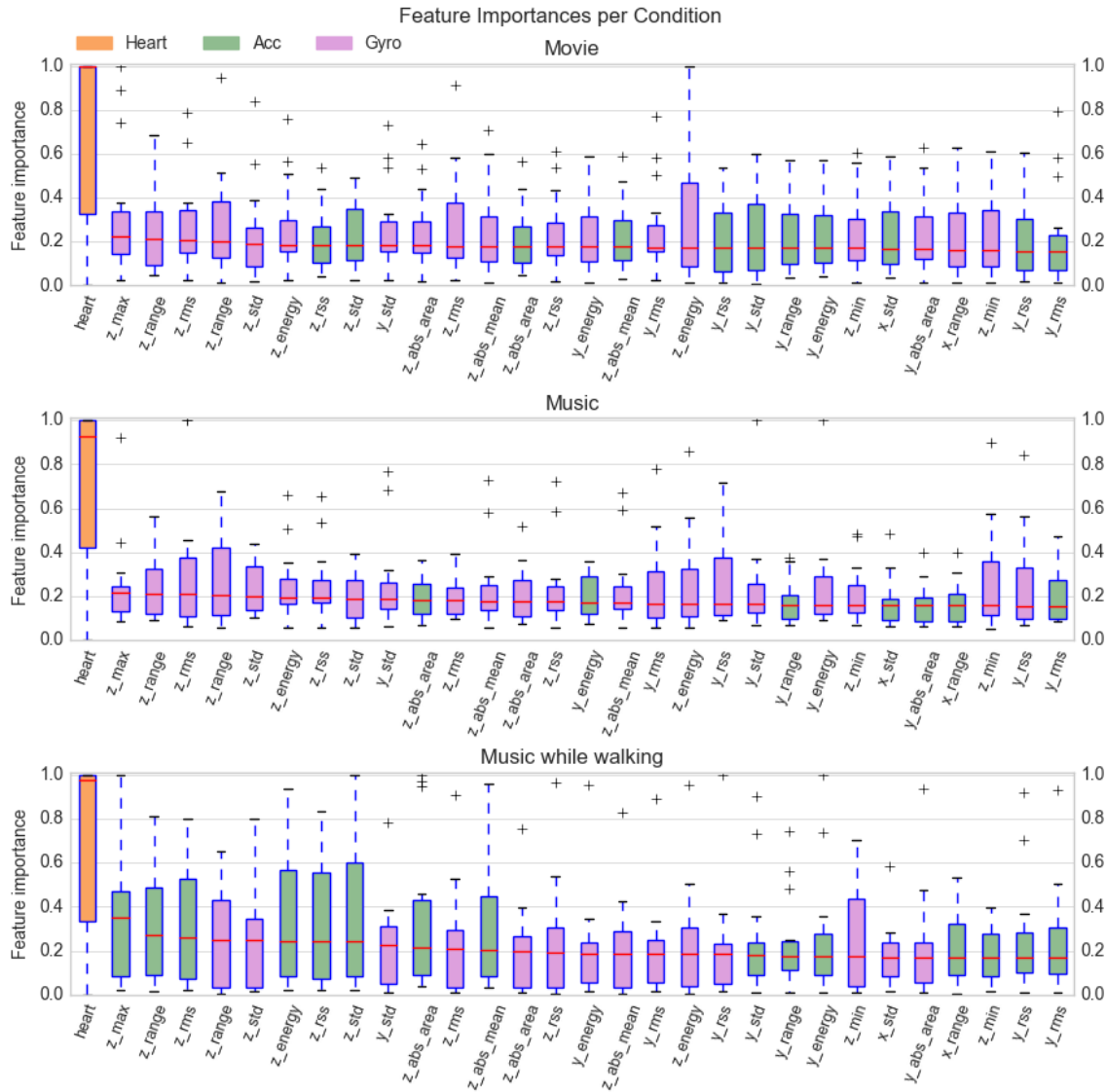

**Figure 6**: Distribution of feature importances per feature for all personal models. The features are colored by the type of sensor from which the feature was extracted. Heart rate was the most important feature for all users in all three conditions. The spread of the boxplot for each feature indicates that the feature was important for some users, but less important for other users.

A compact boxplot indicates that the feature has a similar importance across all users. On the other hand, a boxplot with a large spread indicates that the feature is important for some users, but less important for other users. For all conditions, heart rate was the most important feature. In fact, for Condition 1 (movie) heart rate was the most important feature for at least half of the users (median=1.0). The rest of the features have distributions with smoothly decreasing medians, but with heart rate being the only feature with a clear difference with the other features.

## Discussion

### Behavioral response to stimuli

Participants reported feeling less negative affect after watching sad videos or after listening to the sad music. Unlike the conditions in which listen/watch before walking, participants felt more negative during the happy when compared to neutral when listening to music while walking. Our findings suggest that the walking activity after experiencing a stimulus is useful to alleviate negative mood, similar to (27,28), but not while experiencing the stimuli. One reason for this is that participants were focused on the song and possibly the change between music type creates resentment/unhappiness. Some studies suggest that people may have a preference for sad music (29,30), which may influence the participants' response towards the stimuli. However, a subset of 10 participants reported liking the sad stimulus the least compared to happy and neutral stimuli, suggesting that this is not the case of liking sad music more than others. This personal preference self-report further adds credence to the PANAS results in that walking is useful in alleviating negative mood.

From the heart rate data, our participants did not experience any significant difference in their heart rate between the emotions. This suggests two possible explanations; one being that walking itself is a vigorous activity compared to standing still, thus the brief exposure to emotional stimulus may not have been captured holistically. The other explanation is that both emotions were equally successful in evoking their emotional state hence the non-significant difference between them. Nonetheless the data from PANAS suggest that it is likely the latter because participants reported experiencing the difference between positive and negative states.

### Classifiers for emotion recognition

The high accuracies achieved across all users for the classification of happy vs sad emotional states provides further evidence for the hypothesis that movement sensor data can be used for emotion recognition. To build the personal models, we used statistical features that are computationally cheap, which would make it feasible to deploy an application on a smart watch or a smartphone that can track emotions from movement sensor data without taxing the smart watch or smartphone processor. This means that an application deployed on a smart watch or a smartphone.

Even though the features extracted from gyroscope data and the heart rate increase the overall performance of the models, using only accelerometer data for emotion recognition results in mean accuracies of at least 71% for all conditions. Thus, it is not necessary to collect gyroscope and heart rate data to build high accuracy models. We do note that the APIs of smartphones and smart watches make it just as accessible to retrieve gyroscope and heart rate. In addition, the high importance of the heart rate feature in the random forests models ought to encourage developers to use heart rate data from a smart watch for emotion recognition.

When comparing all of the classification results from the various experiments, we can build high fidelity emotion recognition models for about 25% of the participants, average fidelity models for about 50% of the participants, and low fidelity models for the last 25% of the participants. These results are encouraging, yet they also indicate that further work is needed in order to achieve better performance for about 75% of the participants. For example, this could be achieved by extracting additional features, by using a more complex classifier, or by collecting more data for training and testing personal models. Lastly, our results on emotion cross-validation highlight that the personal models for about half of the participants are learning features that capture emotions.

## Limitations

Previous studies have utilized a contrast experimental paradigm to manipulate participants' moods: positive versus negative mood (2); negative or neutral (31); positive, negative and neutral (32,33) using music or avatars. Past research findings indicate that negative moods tend to reduce mood recovery and a slower response to accurately identify other emotional expressions. While these user studies did not apply to emotion recognition from sensor data from a smart watch, we did not address issues such as reduced mood recovery for participants who were shown the sad stimulus first, although we did perform counterbalancing for our stimuli on our participants.

The integrity of the sensor data is a concern. For Conditions 1 and 2, participants were primed with audio and audio-visual stimulus for a few minutes, but beyond the PANAS scores, we do not have other means to indicate that the stimulus had the intended effect. Furthermore, the effect of the stimulus on the participants is questionable given that participants were not emotionally invested in the movie and music clips that were shown. The personal models do distinguish at high accuracies between features extracted from the happy, sad, and neutral emotions, but we do not know for certain that the happy data is truly associated with a "happy" emotional state in users. In general, given that the mixed-design study consisted of three conditions, 50 participants is a small sample size.

From a modeling and data analysis point of view, the amount of data collected was small. Hence, this limits both the training and validation of the classifiers. While the personal models yielded high accuracies for many users, for other users the results were slightly better than random guessing. Finally, we did not consider more flexible modeling approaches, such as using a time-aware model or using a neural network trained on the raw sensor data, instead of extracting features from sliding windows.

The personal models we built are naïve in that each window is an independent sample. Therefore, a model could potentially predict happy-sad-happy for three consecutive one second windows, which is unrealistic as a user is not likely to go from happy to sad and back to happy in a matter of 3 seconds. This limitation of our modeling approach will be addressed in future work.

### Comparison with Prior Work

Our work is closest to the work in (10,11). In (10), details of the design study are omitted, including the choice of videos and procedure. A limitation in (10) is that data was collected from two smartphones, one strapped to the wrist and one strapped to the ankle of participants. In (11), 123 participants were recruited (twice the size of our sample),  and smart bracelets were used for data collection, with participants wearing a smart bracelet on their wrist and another smart bracelet on their ankle (with the latter violating ecological validity). We achieved accuracies comparable to (11), using only the data from one smart watch on the participants' wrists and without relying on data from other body locations. Our work also differs in that we focus on happy and sad emotional states, whereas in (11) they focused on happy and angry emotional states. In contrast to prior work, we performed more rigorous testing by including emotion cross-validation and by extracting features from accelerometer, gyroscope, and heart rate sensors.

In contrast to emotion prediction based on typing behavior (5,6), smartphone usage (7,8), and smartphone speech recordings (9), we focus on movement data and heart rate data. The EmotionSense system does use accelerometer data to determine whether a user is moving, but not for emotion recognition (9).

### Conclusions and Future Work

Our findings suggest that emotional expression is transparent even in automatic functions such as walking gait. This finding is interesting in that healthy young adults typically do not report large differences in their emotional state, unlike some clinical groups (34).  These findings also validate our methodological approach in priming the emotional state and the subsequent modeling using machine learning algorithms.

Many studies have focused on face and voice modalities, but recent studies have shown that we not only tend to adopt different body postures and gait as a reflection of our emotions, but that they are just as easily recognized by others, indicating that walking gait is a form of social signal. However, the emotional behavioral response is only evident after experiencing the stimulus on its own or while experiencing both together (e.g. listening to music while walking). Nonetheless, our findings add further knowledge in the field of social communication, particularly in specific clinical conditions. The unobtrusive wearable is a good complement for collecting data and providing biofeedback and interventions for emotional regulation. Recent studies have started analyzing the possibility of using wearables to provide more readily available treatment for patients and provide feedback to clinicians to cater for their needs (34–36). The benefits of using these wearables, particularly in identifying emotional states is useful for specific clinical conditions, such as social anxiety and borderline personality disorder. While most research is focused on getting the patients to self-rate their moods, having actigraph data and walking patterns will complement the information necessary for the clinician. Other than for clinical population, this type of information is also useful for vulnerable populations (e.g. older adults) who are experiencing some emotional distress and social isolation (37). Future studies should look into the duration of wearing such wearables (over 24 hours cycles) and duration in experiencing stimuli (acute or chronic experiences).

### Acknowledgements

We thank all the volunteers who participated in our user study. Elisa Roberti and Magdalene Rose for assisting in the data collection.

### Conflicts of Interest

None declared.

### Abbreviations

BL: Baseline
RF: Random Forests
LR: Logistic Regression
API: Application Programming Interface
Acc: Accelerometer
Gyro: Gyroscope
HR: Heart rate